\documentclass[aps,pra,eprint,groupedaddress,showkeys]{revtex4-1}
\usepackage{amssymb,amsmath}
\usepackage{braket,color}
\usepackage{graphicx}
\usepackage{mathrsfs}
\usepackage{bm}
\usepackage{array}
\usepackage{appendix}
\newcommand{\PreserveBackslash}[1]{\let\temp=\\#1\let\\=\temp}
\newcolumntype{C}[1]{>{\PreserveBackslash\centering}p{#1}}

\begin{document}

\title{Fault-tolerant conversion between adjacent Reed-Muller\\ quantum codes based on gauge fixing}
\author{Dong-Xiao Quan$^{1,2}$ }
\email[]{dxquan@xidian.edu.cn}
\author{Li-Li Zhu$^{1}$ }
\author{Chang-Xing Pei$^{1}$ }
\author{Barry C.\ Sanders$^{2,3,4,5}$ }
\affiliation{$^1$State Key Laboratory of Integrated Services Networks, Xidian University, Xi'an, Shaanxi 710071, P.\ R.\ China}
\affiliation{$^2$Institute for Quantum Science and Technology, University of Calgary, Alberta T2N 1N4, Canada}
\affiliation{$^3$Program in Quantum Information Science, Canadian Institute for Advanced Research, Toronto, Ontario M5G 1Z8, Canada}
\affiliation{$^4$Hefei National Laboratory for Physical Sciences at Microscale, University of Science and Technology of China, Hefei, Anhui 230026, P.\ R.\ China}
\affiliation{$^5$Shanghai Branch, CAS Center for Excellence and Synergetic Innovation Center in Quantum Information and Quantum Physics,
      University of Science and Technology of China, Shanghai 201315, P.\ R.\ China}

\date{\today}

\begin{abstract}
We design forward and backward fault-tolerant conversion circuits, which convert between the Steane code and the 15-qubit Reed-Muller quantum code so as to provide a universal transversal gate set.
In our method,
only 7 out of total 14 code stabilizers need to be measured,
and we further enhance the circuit by simplifying some stabilizers;
thus, we need only to
measure eight weight-4
stabilizers for one round of forward conversion and seven weight-4 stabilizers for one round of backward conversion.
For conversion,
we treat random single-qubit errors and their influence on syndromes of gauge operators,
 and our novel single-step process enables
more efficient fault-tolerant conversion between these two codes.
We make our method quite general by showing how to convert between any two adjacent Reed-Muller quantum codes~$\overline{\textsf{RM}}(1,m)$ and~$\overline{\textsf{RM}}\left(1,m+1\right)$,
for which we need only measure stabilizers whose number scales linearly with~$m$
rather than exponentially with~$m$ obtained in previous work.
We provide the explicit
mathematical expression for the necessary stabilizers and the
concomitant resources required.
\end{abstract}
\maketitle

\section{Introduction}
\label{intro}

Quantum information technology, with applications such as secure quantum communication
or universal quantum computing,
is extremely powerful but challenging due to the fragility of quantum information in the presence of noise, loss and decoherence.
Fault-tolerant quantum error correction~\cite{Sho95,Got97,Sho96} ameliorates this problem of fragility by encoding plain-text quantum information into cipher text, processing this encoded information while also measuring error syndromes and correcting, and finally turning back to plain text.
For fault-tolerant error correction,
transversal gates are especially valuable as qubits in each code block act bitwise between corresponding qubits in each code block,
thereby naturally preventing error propagation~\cite{ZCC11}.
Unfortunately, no code can
simply enable a universal set of transversal gates~\cite{ZCC11,EK09,CCC+08} without invoking a technique involving ancillary qubits; thus, complicated strategies are employed to produce universal transversal gate sets.
One strategy to circumvent the non-universal gate set problem
is to employ magic-state distillation~\cite{Bravyi2005,Reichardt2005,Meier2012,Bravyi2012,Jones2013,Fowler2013}.
Although magic-state distillation has the advantage of a higher fault-tolerant threshold,
the overhead in preparation and distillation is a major bottleneck for scalable quantum computing~\cite{Jochym-OConnor2014,SR-2014,Ducloscianci2015,Beverland2016}.

Alternative techniques can provide a universal fault-tolerant gate set.
The Steane code~\cite{Steane1996} provides transversal Clifford gates but not the transversal
$\text{T}:=\operatorname{diag}(1,\exp\{\text{i}\pi/4\})$
gate whereas the 15-qubit Reed-Muller quantum code (RMQC)~\cite{Steane1996b} is transversal for
\textsc{T}, controlled-NOT (\textsc{CNOT}), controlled-S (\textsc{CS}), controlled-Z (\textsc{CZ}), and
controlled-controlled-\textsc{Z} (\textsc{CCZ}) gates,
but not for the Hadamard (\textsc{H}) gate.
One approach to achieving a universal set of gates
employs just one code such as the Steane code or the 15-qubit RMQC
and does not invoke code conversion.
For example, for the Steane code, the fault-tolerant \textsc{T} gate can be realized with the help of an ancillary logical qubit, Pauli Z-measurement and transversal $S:=\operatorname{diag}(1,\text{i})$ and \textsc{X} operations, which we call the standard method to construct T gate for the Steane code~\cite{Nielsen2000}.
Alternatively,
for the 15-qubit RMQC,
which does not include the transversal \textsc{H} gate in the universal gate set,
a Hadamard gadget can be introduced:
this Hadamard gadget exploits the transversality of the CS gate,
which yields a \textsc{CZ} gate if applied twice,
and, combined with an ancillary logical qubit and a Pauli X-measurement and post-selection,
yields a Hadamard gate~\cite{Bremner2011}. A universal gate set for 15-qubit RMQC can also be achieved by
using transversal operations and the gauge-fixing method to realize the fault-tolerant \textsc{H} gates~\cite{Paetznick2013,Bombin2013,Kubica2015}.

Another technique is code concatenation,
for which the qubits that make up the code are subsequently encoded into a second code so combining the 7-qubit Steane code and 15-qubit RMQC requires 105 qubits~\cite{Jochym-OConnor2014,Jochym-OConnor2016}.

An alternative method is by code conversion~\cite{Anderson2014,Hwang2015,Bravyi2015,Hill2013,Choi2015,Bombin2016}, which converts between two codes to provide a universal transversal gate set. Direct conversion can be realized via Clifford operations, which is general for converting between different kinds of codes but requires 13 single-qubit and 74 two-qubit gates. Furthermore, this direct conversion requires error correction at every step to ensure fault tolerance~\cite{Hwang2015}. Another direct conversion method is by gauge fixing as both codes correspond to the same subsystem code with different gauge qubits~\cite{Anderson2014,Bravyi2015}, which is more efficient compared to the method of Clifford operations with respect to the number of qubits and gates required. This method is also studied from the view of colour codes~\cite{Bravyi2015} and furthermore generalized to the dimensional jump between  two-dimensional  and  three-dimensional colour codes or even higher dimensions~\cite{Bombin2016}. This code conversion can also be realized by code teleportation~\cite{Choi2015,Choi2013}.

As both the Steane code and the RMQC are of the same code distance $3$, they can only correct one error.
For fixed distance,
fewer qubits are better as fewer errors can happen
so smaller codes are easier to implement and need fewer resources to perform syndrome measurements.
Thus, the Steane code is more efficient than the 15-qubit RMQC.
Detailed analysis shows that the overhead of using only the Steane code to realize
a fault-tolerant \textsc{T} gate is lower compared with direct conversion~\cite{Anderson2014}
and with the teleportation scheme~\cite{Choi2015}, since implementation of the \textsc{T} gate is feasible in the Steane code with ancilla,
and the necessary stabilizer measurements in the direct conversion method require many gates~\cite{Choi2015,Choi2013}.
In this paper we find that some stabilizers are not necessary for fault-tolerant conversion
between adjacent RMQCs based on gauge fixing, especially for higher $m$,
so that we can significantly decrease resource requirement.

Here we significantly
improve the fault-tolerant code-conversion method of Anderson, Duclos-Cianci and Poulin (ADP14)~\cite{Anderson2014} by
exponentially
 reducing the number of stabilizer measurements, which thus reduces the resource requirement. Our circuits and method have the following advantages. First we split some of the stabilizers into two gauge operators, which reduces the resource requirement for the measurements of the syndromes. Second in the single-step process we consider random single-qubit errors and their influence on the syndromes of the gauge operators, using fixed syndromes to choose the operations so that we can make fault-tolerant conversion. Third we generalize the method to any adjacent RMQCs for which we need only measure $2m+1$ stabilizers of the RMQC code.
In contrast ADP14's approach discusses measuring all $2^{m+1}-2$ stabilizers
and requiring $2^{m+1}-m-2$ stabilizers.
Our approach significantly reduces this stabilizer overhead.

The rest of this paper is organized as follows.
In \S\ref{sec:basicknowledge}, we first give a brief review of the order-one RMQC and then give the stabilizer groups for the 7-qubit Steane code, the 15-qubit RMQC
and
extended Reed-Muller quantum code (ERMQC) and,  in \S\ref{sec:ADP14} we describe the ADP14 method. In \S\ref{sec:convertSteaneRMQC} we describe forward conversion from the 7-qubit Steane code to the 15-qubit RMQC in detail, and we deal with backward conversion in \S\ref{sec:convertRMQCSteane}.
In \S\ref{sec:convertmtomplus1}, we explain how to convert between adjacent order-one RMQCs and, in \S\ref{sec:simulationandcost} we elaborate on our simulation process and cost analysis. We conclude in \S\ref{sec:conclusion}.

\section{Basic Knowledge}
\label{sec:basicknowledge}
In this section, we give the basic concept of quantum error correction (QEC) and the stabilizer groups for the Steane code, the 15-qubit RMQC and ERMQC.
From the relations for the stabilizer groups, we design the conversion scheme in following sections.

QEC~\cite{Sho95,Got97,Sho96} is an effective way to protect the information from the influence of noise. The quantum block notation $[[n,k,d]]$ refers to encoding~$k$ logical qubits into~$n$ physical qubits with code distance~$d$ so that $t<d/2$ random errors can be detected and corrected by syndrome measurements and error correction processes.
The order-one RMQC~\cite{Steane1996b}
is written as $\overline{\textsf{RM}}(1,m)=[[M-1,1,3]]$, which encodes 1 logical qubit into $M-1$ qubits with distance 3 for $M:=2^m$. Adjacent codes are~$\overline{\textsf{RM}}(1,m)$ and~$\overline{\textsf{RM}}\left(1,m+1\right)$. The Steane code is written as $\overline{\textsf{RM}}(1,3)$ and the 15-qubit RMQC is written as $\overline{\textsf{RM}}(1,4)$; in the framework of RMQC, they are adjacent RMQCs.

RMQCs are derived from the recursively defined classical order-one Reed-Muller code~\cite{FN1977}. By deleting the first row and column of the generator matrix for the classical Reed-Muller code, we obtain $\overline{\textsf{RM}}(1,3)$'s generator matrix
\begin{align}
\label{eq:GSR}
    \bar{G}(1,3)
    :=\begin{bmatrix}
    \bar{G}(1,3)_1\\
    \bar{G}(1,3)_2\\
    \bar{G}(1,3)_3\\
    \end{bmatrix}
    =\begin{bmatrix}
               1&0&1&0&1&0&1\\
               0&1&1&0&0&1&1\\
               0&0&0&1&1&1&1\\
               \end{bmatrix},
\end{align}
which can be used recursively to obtain
the generator matrix for $\overline{\textsf{RM}}\left(1,m+1\right)$
given by
\begin{align}
\label{eq:Gm1}
    \bar{G}\left(1,m+1\right)=\begin{bmatrix}
              \bar{G}(1,m)& 0 &\bar{G}(1,m)\\
               \bm{\bar{0}}_{M-1}& 1 & \bm{\bar{1}}_{M-1}\\
              \end{bmatrix},\
 	\bar{G}(1,m)
		\!=\!\begin{bmatrix}
			\bar{G}(1,m)_1\\
				\vdots\\
			\bar{G}(1,m)_m\\
 		\end{bmatrix},
\end{align}
where
\begin{equation}
	\bm{\bar{b}}_{M}
		:=b^{\otimes{M}},\;
		b\in\{0,1\}.
\end{equation}

As all RMQCs are Calderbank-Shor-Steane codes~\cite{Steane1996a,Calderbank1996},
we can subdivide their stabilizers  into two parts
corresponding to phase errors ($X$ stabilizers) and flip errors ($Z$ stabilizers).
The $X$ parts are obtained from the generator matrix of the code, and the $Z$ parts are obtained from the generator matrix of the dual code, which is also the parity-check matrix of the code.
As $\overline{\textsf{RM}}(1,3)$ is self dual~\cite{FN1977}, the stabilizers of the Steane code can be written as
\begin{equation}
\bm{S}(1,3)=\left\langle \bar{G}(1,3)^X, \bar{G}(1,3)^Z\right\rangle,
\end{equation}
with
\begin{align}
\label{eq:G13}
	\bar{G}(1,3)_1^{B\in \{X,Z\}}
		=&B_1B_3B_5B_7,
			\nonumber\\
	\bar{G}(1,3)_2^{B\in \{X,Z\}}
		=&B_2B_3B_6B_7,
			\nonumber\\
	\bar{G}(1,3)_3^{B\in \{X,Z\}}
		=&B_4B_5B_6B_7,
\end{align}
where~$G^{B\in\{X,Z\}}$ is a matrix obtained from $G$ by substituting 1 by~$B$
and substituting 0 by~$I$.
As the Steane code is self dual, it is transversal for logical
\begin{equation}
\label{eq:barHn}
	\bar{H}:=H^{\otimes{n}},
\end{equation}
where $n$ is the number of qubits for the code~\cite{Bravyi2015,Betsumiya2012}.

When $m\geq3$, $\overline{\textsf{RM}}(1,m)$ is contained in its dual~\cite{FN1977}. For the case $m=4$, we define a matrix
\begin{equation}
\label{eq:H14}
\tilde{H}(1,4):= \begin{bmatrix}
    \tilde{H}(1,4)_1\\
   \tilde{H}(1,4)_2\\
    \tilde{H}(1,4)_3\\
    \tilde{H}(1,4)_4\\
    \tilde{H}(1,4)_5\\
    \tilde{H}(1,4)_6\\
    \end{bmatrix}
=\begin{bmatrix}
               1&0&1&0&0&0&0&0&1&0&1&0&0&0&0\\
               0&1&1&0&0&0&0&0&0&1&1&0&0&0&0\\
               0&0&1&0&0&0&1&0&0&0&1&0&0&0&1\\
                \multicolumn{7}{c}{\bar{G}_{1,3}}& \multicolumn{8}{c}{\bm{\bar{0}}_{8}} \\
               \end{bmatrix},
\end{equation}
and we have
\begin{align}
\label{eq:H14}
	\tilde{H}(1,4)_1^{B\in \{X,Z\}}
		=&B_1B_3B_9B_{11},\nonumber\\
	\tilde{H}(1,4)_2^{B\in \{X,Z\}}
		=&B_2B_3B_{10}B_{11},\nonumber\\
	\tilde{H}(1,4)_3^{B\in \{X,Z\}}
		=&B_3B_7B_{11}B_{15},
\end{align}
with~$\tilde{H}$ and~$\tilde{H}^{B\in \{X,Z\}}$ related in the same way as~$G$ and~$G^{B\in \{X,Z\}}$ earlier.

Obviously
\begin{equation}
	\tilde{H}(1,4)\bar{G}(1,4)^T=0
\end{equation}
and the parity-check matrix for $\overline{\textsf{RM}}(1,4)$
is the vertical concatenation of the two matrices
$\bar{G}(1,4)$ and~$\tilde{H}(1,4)$.
Thus, the stabilizers of the 15-qubit RMQC are
\begin{equation}
\label{eq:SRM}
\bm{S}(1,4)=\left\langle \bar{G}(1,4)^X, \bar{G}(1,4)^Z, \tilde{H}(1,4)^Z\right\rangle
\end{equation}
with
\begin{align}
\label{eq:G14}
	\bar{G}(1,4)_1^{B\in \{X,Z\}}
		=&B_1B_3B_5B_7B_9B_{11}B_{13}B_{15},
			\nonumber\\
	\bar{G}(1,4)_2^{B\in \{X,Z\}}
		=&B_2B_3B_6B_7B_{10}B_{11}B_{14}B_{15},
			\nonumber\\
	\bar{G}(1,4)_3^{B\in \{X,Z\}}
		=&B_4B_5B_6B_7B_{12}B_{13}B_{14}B_{15},
			\nonumber\\
	\bar{G}(1,4)_4^{B\in \{X,Z\}}
		=&B_8B_9B_{10}B_{11}B_{12}B_{13}B_{14}B_{15}.
\end{align}

As the syndromes of $\bar{G}(1,4)^Z$ can unambiguously discriminate all single-qubit~$X$ errors,
the syndromes of $\tilde{H}(1,4)^Z$ are often omitted. The code is triply-even as its weight is 8 so it is transversal for logical $\overline{T}:=T^{\otimes{n}}$~\cite{Bravyi2015,Betsumiya2012}.

From Eq.~(\ref{eq:Gm1}) we can see the 15-qubit RMQC comprises two blocks of Steane code for the first and last 7-qubit blocks
plus another interconnecting qubit labeled~8.
This interconnecting qubit entangles with the last 7-qubit block of Steane code.
Thus, we can prepare an 8-qubit quantum state $\ket{\phi}=1/\sqrt{2}\left(\ket{0}\ket{\bar{0}}_{3}+\ket{1}\ket{\bar{1}}_{3}\right)$ in the last 8 qubits~\cite{Anderson2014},
where $\ket{\bar{b}}_{m},\;b\in\{0,1\}$,
is the logical bit~$b$ encoded into the~$\overline{\textsf{RM}}(1,m)$ code.

Together with the Steane code $\ket{\psi}=\alpha\ket{\bar{0}}_{3}+\beta\ket{\bar{1}}_{3}$,
we construct a 15-qubit ERMQC
\begin{equation}
\label{eq:noerrorcase}
	\ket{\phi}_{3\backsim4}
		=1/\sqrt{2}\left(\alpha\ket{\bar{0}}_{3}+\beta\ket{\bar{1}}_{3}\right)\left(\ket{0}\ket{\bar{0}}_{3}+\ket{1}\ket{\bar{1}}_{3}\right).
\end{equation}
We refer to Eq.~(\ref{eq:noerrorcase})
as the no-error case for preparation of the 15-qubit ERMQC,
and this ideal input preparation~(\ref{eq:noerrorcase})
is justified in the realistic case by introducing
fault-tolerant preparation using gadgets~\cite{Aliferis2006}.
The stabilizer group for this code can be expressed as
 \begin{equation}
 \label{eq:SEXRM}
	\bm{S}_{3\backsim4}
		=\left\langle\left(\bar{G}(1,3)\otimes\bm{\bar{0}}_{8}\right)^{X},\bar{G}(1,4)^X, \left(\bar{G}(1,3)\otimes\bm{\bar{0}}_{8}\right)^{Z}, \bar{G}(1,4)^Z\right\rangle.
\end{equation}
It is also transversal for logical~$\bar{H}$ as it is self dual. There is no information in the last 8 qubits so it has the same logical operations as the Steane code.

We define a subsystem code with the stabilizers of
\begin{align}
	\bm{S}^\text{sub}_{3\backsim4}
		=\left\langle\bar{G}(1,4)^X,\left(\bar{G}(1,3)\otimes\bm{\bar{0}}_{8}\right)^{Z},\bar{G}(1,4)^Z\right\rangle.
\end{align}
Pauli operators that commute with both the stabilizers and logical operators generate the gauge group~\cite{Bravyi2015} so the gauge group of this subsystem code is
\begin{align}
\bm{G}^\text{sub}_{3\backsim4}=\left\langle \bar{G}(1,4)^X,
\left(\bar{G}(1,3)\otimes\bm{\bar{0}}_{8}\right)^{X}, \bar{G}(1,4)^Z, \tilde{H}(1,4)^Z\right\rangle.
\end{align}
Both the 15-qubit RMQC and the ERMQC belong to this subsystem code with different gauge operators so they can convert to each other by gauge fixing.

\section{ADP14}
\label{sec:ADP14}

Based on the theory of subsystem codes,
ADP14 shows how to convert from $\overline{\textsf{RM}}(1,m)$ to $\overline{\textsf{RM}}(1,{m+1})$.
They first fault-tolerantly prepare the ERMQC, fault-tolerantly measure the $2^{m+1}-2$ stabilizer generators of $\overline{\textsf{RM}}(1,{m+1})$, error-correct given the first $2^{m+1}-m-2$ syndrome bits, and restore the last~$m$ syndrome bits using their associated pure errors.
To convert from $\overline{\textsf{RM}}(1,m+1)$ to $\overline{\textsf{RM}}(1,m)$, they simply fault-tolerantly measure the $2^{m+1}-2$ stabilizer generators of ERMQC, use the first $2^{m+1}-m-2$ syndrome bits to diagnose errors, and restore the last~$m$ syndrome bits using the associated pure errors~\cite{Anderson2014}.

Here we improve the fault-tolerant code conversion method of ADP14 by exponentially reducing the number of stabilizer measurements from $2^{m+1}-2$ to $2m+1$,
which thereby significantly reduces the resource requirement.
Furthermore we use the relations between these stabilizers to simplify the stabilizer measurements.
For code conversion,
we consider single-qubit errors and their influence on the gauge operators so as to achieve fault-tolerant conversion. In the following we first use conversion between the Steane code and the 15-qubit RMQC as an example to describe conversion,
simplification of stabilizer measurements, the single-step error-correction and code-conversion process, and then generalize this conversion to any adjacent RMQCs.

\section{Conversion from the Steane code to the 15-qubit RMQC}
\label{sec:convertSteaneRMQC}
In this section we deal with forward conversion,
which converts from the Steane code to the 15-qubit RMQC.
From Eqs.~(\ref{eq:SRM}) and~(\ref{eq:SEXRM}),
we see that the ERMQC satisfies all the other stabilizers of the 15-qubit RMQC except for $\tilde{H}(1,4)_i^Z,i=1,2,3$.
Our procedure entails measuring
the stabilizers of $\tilde{H}(1,4)_i^Z$, $i=1,2,3$, given in Eq.~(\ref{eq:H14}).
The $i^\text{th}$ $Z$ syndrome is
\begin{equation}
\label{eq:SiMHi}
	S_i=\mathbb{M}\left(\tilde{H}(1,4)_i^Z\right)\in\{0,1\}, ~~i=1,2,3,
\end{equation}
for~$\mathbb{M}(\tilde{H})$
denoting the syndrome measurement result of stabilizer $\tilde{H}$.
Operations are then performed on the collapsed state according to syndromes
as shown in Table~\ref{table:syndrome} in order to get the 15-qubit RMQC.
\begin{table}
\caption{Relation between stabilizer measurement results and corresponding operations for forward conversion in the no-error case~(\ref{eq:noerrorcase})
with $S_i=\mathbb{M}\left(\tilde{H}(1,4)_i^Z\right)$.}
\label{table:syndrome}
\begin{tabular}{C{1.2cm}C{1.2cm}C{1.2cm}C{3.5cm}C{1.2cm}C{1.2cm}C{1.2cm}C{3.5cm}}
\hline\noalign{\smallskip}
$S_1$ & $S_2$ & $S_3$ & operation & $S_1$ & $S_2$ & $S_3$ & operation\\
\noalign{\smallskip}\hline\noalign{\smallskip}
0 & 0 & 0 & I & 1 & 0 & 0 & $X_{10}X_{11}X_{14}X_{15}$\\
0 & 0 & 1 & $X_{12}X_{13}X_{14}X_{15}$ & 1 & 0 & 1 & $X_{10}X_{11}X_{12}X_{13}$\\
0 & 1 & 0 & $X_{9}X_{11}X_{13}X_{15}$ & 1 & 1 & 0 & $X_{9}X_{10}X_{13}X_{14}$\\
0 & 1 & 1 & $X_{9}X_{11}X_{12}X_{14}$ & 1 & 1 & 1 & $X_{9}X_{10}X_{12}X_{15}$\\
\hline\noalign{\smallskip}
\end{tabular}
\end{table}

The operations shown in Table~\ref{table:syndrome} are chosen as follows.
For example, if we obtain $S_1=S_2=0$ and $S_3=1$,
then we need to use~$X$-type operations, which commute with all $Z$-type stabilizers except $\tilde{H}(1,4)_3^Z$ so we can choose $X_{12}X_{13}X_{14}X_{15}$ as the operations. If we obtain $S_1=S_2=1$ and $S_3=0$, then we can choose the operations as $X_{9}X_{10}X_{13}X_{14}$, which anti-commute with $\tilde{H}(1,4)_1^Z$ and $\tilde{H}(1,4)_2^Z$ but commute with all the others.

We can divide our analysis into two cases:
the no-error case, which assumes ideal state preparation~(\ref{eq:noerrorcase}),
and the case of a single--qubit error occurring before the code conversion.
In the case of a single-qubit error,
this error can influence the syndromes of Table~\ref{table:syndrome} and accordingly result in wrong operations
so we need to distinguish single-qubit errors and fix the syndromes before the corrections in Table~\ref{table:syndrome}.
The syndromes of $S(1,4)$ are
\begin{equation}
\label{eq:S14iB}
	S(1,4)_i^{B\in\{X,Z\}}
		=\mathbb{M}\left(\bar{G}(1,4)_i^{B\in\{X,Z\}}\right), ~~i=1,2,3,4,
\end{equation}
and the single-qubit errors are shown in Table~\ref{table:errorsyndrome}.
\begin{table}
\caption{Relation between stabilizer measurement results and single-qubit errors.
The syndromes indicate the bit error when $B=Z$, and indicate the phase error when $B=X$.}
\label{table:errorsyndrome}
\begin{tabular}{C{1.5cm}C{1.5cm}C{1.5cm}C{1.5cm}C{1.5cm}C{1.5cm}C{1.5cm}C{1.5cm}C{1.5cm}C{1.5cm}}
\hline\noalign{\smallskip}
$S(1,4)_1^B$ & $S(1,4)_2^B$&$S(1,4)_3^B$ & $S(1,4)_4^B$ & error & $S(1,4)_1^B$ & $S(1,4)_2^B$ & $S(1,4)_3^B$ & $S(1,4)_4^B$ & error\\
\noalign{\smallskip}\hline\noalign{\smallskip}
0 & 0 & 0 & 0 & NO & 0 & 0 & 0 & 1 & 8\\
0 & 0 & 1 & 0 & 4  & 0 & 0 & 1 & 1 & 12\\
0 & 1 & 0 & 0 & 2  & 0 & 1 & 0 & 1 & 10\\
0 & 1 & 1 & 0 & 6 &0 & 1 & 1 & 1 & 14\\
1 & 0 & 0 & 0 & 1 & 1 & 0 & 0 & 1 & 9\\
1 & 0 & 1 & 0 & 5  & 1 & 0 & 1 & 1 & 13\\
1 & 1 & 0 & 0 & 3  & 1 & 1 & 0 & 1 & 11\\
1 & 1 & 1 & 0 & 7  &1 & 1 & 1 & 1 & 15\\
\noalign{\smallskip}\hline
\end{tabular}
\end{table}

By the~$X$-type syndromes of Table~\ref{table:errorsyndrome}, we can identify the single-qubit~$Z$ error represented as $Z_i$.
The error $Z_i$ has no influence on the syndromes of Table~\ref{table:syndrome} so we need only add the $Z_i$ operation in the correction process.
Similarly, by the~$Z$-type syndromes of Table~\ref{table:errorsyndrome}, we can identify the single-qubit~$X$ error written as $X_j$.
However, this~$X$ error could also result in wrong syndromes of Table~\ref{table:syndrome}
so we need to consider the influence of this error when choosing the operations.

\begin{figure}
\begin{center}
 \includegraphics[width=0.8\textwidth]{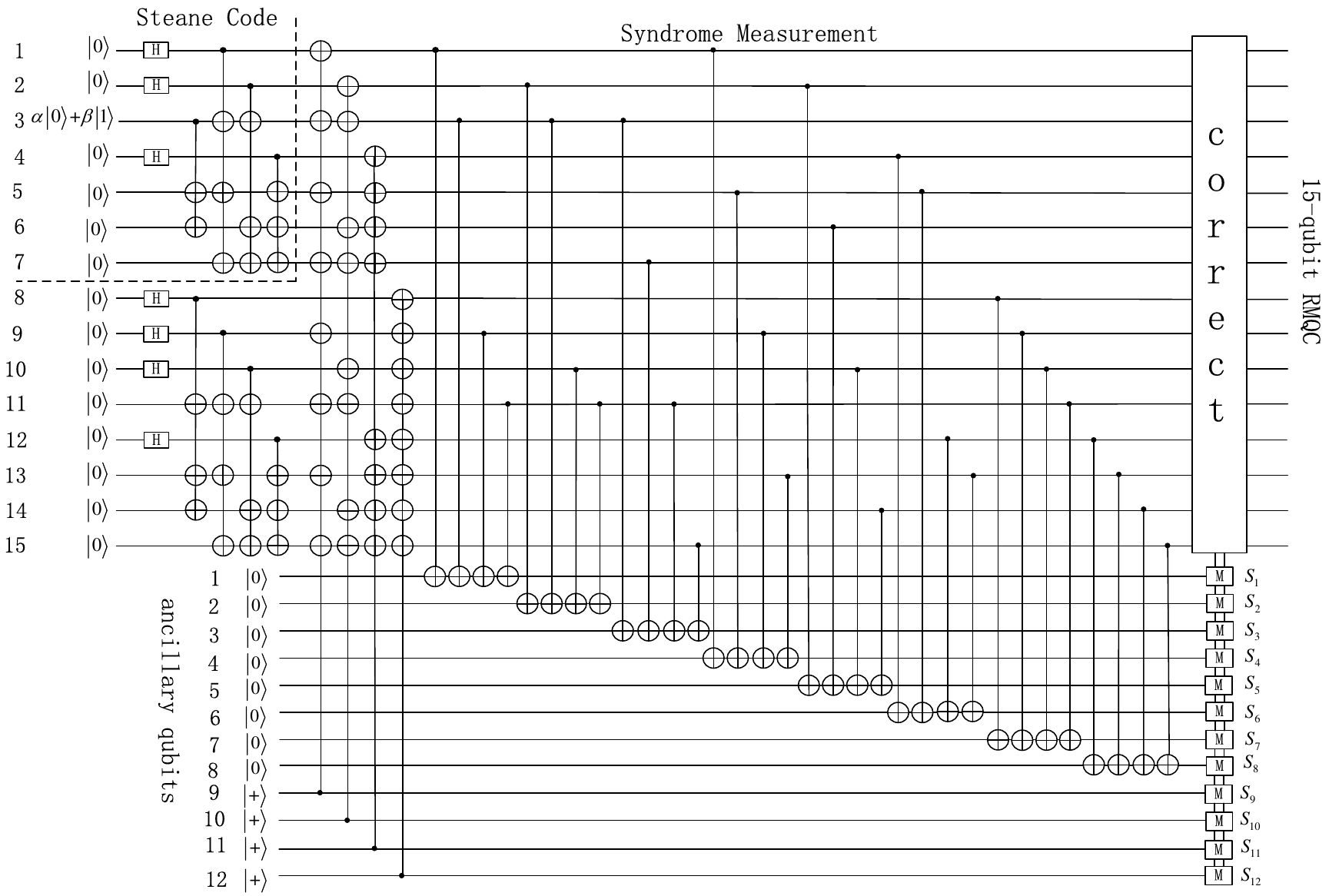}
\caption{%
Conversion circuit from Steane code to 15-qubit RMQC.
Quantum information is first encoded into the Steane code
 $\ket{\psi}=\alpha\ket{\bar{0}}_{3}+\beta\ket{\bar{1}}_{3}$ in qubits 1--7.
At the same time qubits 8--15 are prepared in the state
$\ket{\phi}=1/\sqrt{2}\left(\ket{0}\ket{\bar{0}}_{3}+\ket{1}\ket{\bar{1}}_{3}\right)$.
The ancillary qubits 1--12 are used to measure the syndromes
$S_1=\mathbb{M}\left(Z_1Z_3Z_9Z_{11}\right)$,
$S_2=\mathbb{M}\left(Z_2Z_3Z_{10}Z_{11}\right)$,
$S_3=\mathbb{M}\left(Z_3Z_7Z_{11}Z_{15}\right)$,
$S_4=\mathbb{M}\left(Z_1Z_5Z_{9}Z_{13}\right)$,
$S_5=\mathbb{M}\left(Z_2Z_6Z_{10}Z_{14}\right)$,
$S_6=\mathbb{M}\left(Z_4Z_5Z_{12}Z_{13}\right)$,
$S_7=\mathbb{M}\left(Z_8Z_{9}Z_{10}Z_{11}\right)$,
$S_8=\mathbb{M}\left(Z_{12}Z_{13}Z_{14}Z_{15}\right)$,
$S_9=\mathbb{M}\left(X_1X_3X_5X_7X_9X_{11}X_{13}X_{15}\right)$,
$S_{10}=\mathbb{M}\left(X_2X_3X_6X_7X_{10}X_{11}X_{14}X_{15}\right)$,
$S_{11}=\mathbb{M}\left(X_4X_5X_6X_7X_{12}X_{13}X_{14}X_{15}\right)$,
$S_{12}=\mathbb{M}\left(X_8X_{9}X_{10}X_{11}X_{12}X_{13}X_{14}X_{15}\right)$.}
\label{fig:forwardconversion}
\end{center}
\end{figure}

Let us consider the point in Fig.~\ref{fig:forwardconversion} subsequent to the $S_3$ measurement
at which point we have measured all three syndromes $S_1$, $S_2$ and $S_3$.
Then we can simplify the measurement according to
\begin{align}
\label{eq:S141ZM}
	S\left(1,4\right)_1^Z
		=&\mathbb{M}\left(\bar{G}(1,4)_1^{Z}\right)
			\nonumber\\
		=&\mathbb{M}\left(Z_1Z_3Z_5Z_7Z_9Z_{11}Z_{13}Z_{15}\right)
			\nonumber\\
		=&\mathbb{M}\left(Z_3Z_7Z_{11}Z_{15}Z_1Z_5Z_9Z_{13}\right)
			\nonumber\\
		=&\underbrace{\mathbb{M}\left(Z_3Z_7Z_{11}Z_{15}\right)}_{S_3}
			\oplus\underbrace{\mathbb{M}\left(Z_1Z_5Z_9Z_{13}\right)}_{S_4}
			\nonumber\\
		=&S_3\oplus S_4,
\end{align}
where we have used commutativity of~$Z$ operators in the second step
and, for the third step, we have used the syndrome definition~(\ref{eq:SiMHi})
and the expression for~$\tilde{H}(1,4)_i^Z$ given in Eq.~(\ref{eq:H14}).
Generalizing Eq.~(\ref{eq:S141ZM}),
we obtain the simplifying assignments
\begin{align}
	\mathbb{M}\left(\bar{G}(1,4)_1^{Z}\right)\to&S_4=\mathbb{M}\left(Z_1Z_5Z_{9}Z_{13}\right),
		\nonumber\\
	\mathbb{M}\left(\bar{G}(1,4)_2^{Z}\right)\to&S_5=\mathbb{M}\left(Z_2Z_6Z_{10}Z_{14}\right),
		\nonumber\\
	\mathbb{M}\left(\bar{G}(1,4)_3^{Z}\right)\to&S_6=\mathbb{M}\left(Z_4Z_5Z_{12}Z_{13}\right).
\end{align}
Unfortunately
\begin{equation}
	\mathbb{M}\left(\bar{G}(1,4)_4^{Z}\right)=\mathbb{M}\left(Z_8Z_{9}Z_{10}Z_{11}Z_{12}Z_{13}Z_{14}Z_{15}\right)
\end{equation}
cannot be simplified in this way.
However, it can be divided into two weight-4 stabilizers
\begin{align}
	S_7=&\mathbb{M}\left(Z_8Z_9Z_{10}Z_{11}\right),
		\nonumber\\
	S_8=&\mathbb{M}\left(Z_{12}Z_{13}Z_{14}Z_{15}\right).
\end{align}
 in order to be experimentally realistic since both of them commute with all the stabilizers.

The total set of~$Z$-type syndromes is $\{S_1,S_2,\dots,S_8\}$.
Using the technique shown in Eq.~(\ref{eq:S141ZM}),
we obtain three additional relations
\begin{equation}
\label{eq:S1423ZM}
	S(1,4)_2^Z=S_3\oplus S_5,\;
	S(1,4)_3^Z=S_3\oplus S_5\oplus S_2\oplus S_6,\;
    S(1,4)_4^Z=S_7\oplus S_8
\end{equation}
and Table~\ref{table:errorsyndrome} to deduce the single-qubit~$X$ error $X_j$.
Then we modify the first three syndromes~$S_{1,2,3}$ using the relation
\begin{equation}
\label{eq:syndromecorrection}
	Z_j\in \tilde{H}(1,4)_n^Z\implies S_n=S_n\oplus1
	(n=1,2,3).
\end{equation}
Now we use the modified $S_{1,2,3}$ and Table~\ref{table:syndrome} to deduce the operations written as~$X_\text{sub}$. Thus, the total operations we need to perform for the correction process are
$X_\text{sub}X_jZ_i$; by these operations we obtain the 15-qubit RMQC.

The forward conversion circuit is shown in Fig.~\ref{fig:forwardconversion}, where we first prepare the entangled state of a single qubit with the Steane-code qubits, these ancillary qubits together with the initial Steane code form the ERMQC.
Then we measure four weight-8 X type stabilizers and eight weight-4 Z type stabilizers.
Preparation and syndrome measurements shown in Fig.~\ref{fig:forwardconversion} are not fault-tolerant as the single error may spread into multiple errors. The fault-tolerant preparation can be realized by using gadgets~\cite{Aliferis2006}.
In order to ensure fault-tolerant measurement,
we use the Shor state~\cite{Weinstein2015},
or use encoded Bell pairs~\cite{Knill2005}, use encoded blocks~\cite{Steane1997,Weinstein2016}, or by adding flags~\cite{Chao2017} which reduces requisite resources significantly, to perform syndrome measurement, repeat the syndrome measurement three times and then make a majority vote to ensure fault-tolerant measurement.

Recall that we separately consider the no-error case~(\ref{eq:noerrorcase})
and the case of a single-qubit error.
Even in the no-error case,
subsequent to the $S_1$, $S_2$ and $S_3$ syndrome measurements,
the ERMQC collapses to the 15-qubit RMQC with probability~$1/8$.
Thus, we need to apply the fixing operation shown in Table~\ref{table:syndrome} with probability~$7/8$.
This fixing operation is shown as the correct box in Fig.~\ref{fig:forwardconversion}.
After this correction we need another round of syndrome measurements;
this round needs~$8$ ancillary qubits to measure the stabilizers of~$\bar{G}(1,4)^X$ and~$\bar{G}(1,4)^Z$
in order to judge whether an error happened during the correction process.
If an error happened in the correction phase,
then we need to repeat the correction and syndrome measurement one more time.
As we assume the error probability is sufficiently small that we can ignore more than one error occurring during the entire process,
we are sure that all eight syndromes are zero in this final step, which we verify by this final syndrome measurement.

If we only want to fault-tolerantly convert from the Steane code to the 15-qubit RMQC, as the code distance is 3, one error is allowed in the final code block.
Single Z errors do not have any influence on the gauge operators,
summarized in Table~\ref{table:syndrome},
that we have used to do conversion, so these single Z errors do not propagate to multiple errors. Thus, we do not need to measure the four X type syndromes which are used to diagnose the Z errors, the ancillary qubits 9-12 in Fig.~\ref{fig:forwardconversion} can be omitted, this time we need only measure eight weight-4 Z type stabilizers. After the correct operation, there may be one single-qubit error in the code, but fault-tolerant conversion is fulfilled.

\section{Conversion from the 15-qubit RMQC to the Steane code}
\label{sec:convertRMQCSteane}
In this section we deal with the backward conversion, which first converts from the 15-qubit RMQC to the ERMQC£¬
then drops the last 8 qubits to get the Steane code.

As in \S\ref{sec:convertSteaneRMQC}, we first deal with the ideal case of no error in the 15-qubit RMQC.
From Eqs.~(\ref{eq:SRM}) and~(\ref{eq:SEXRM}),
we see that the 15-qubit RMQC satisfies all the other stabilizers of the ERMQC except for $\bar{G}(1,3)_i^X$, $i=1,2,3$.
Thus, we measure these three stabilizers $\bar{G}(1,3)_i^X$, $i=1,2,3$, given in Eq.~(\ref{eq:G13}), and write
the $i^\text{th}$ $X$ syndrome as
\begin{equation}
\label{eq:SiMGi}
	S'_i=\mathbb{M}\left(\bar{G}(1,3)_i^X\right)\in\{0,1\},~~i=1,2,3.
\end{equation}
Following the syndrome measurements,
we perform the fixing operations on the collapsed state according to Table~\ref{table:backwardsyndrome}.
In Table~\ref{table:backwardsyndrome} the operations are chosen similar as before in \S\ref{sec:convertSteaneRMQC}.
For example, if we obtain $S'_1=S'_2=0$ and $S'_3=1$,
then we can choose $Z_{3}Z_{7}Z_{11}Z_{15}$ as the fixing operations, which anti-commute with $\bar{G}(1,3)_3^X$ but commute with all the other stabilizers of $\bm{S}_{3\backsim4}$.

\begin{table}
\caption{Relation between stabilizer measurement results and corresponding operations for backward conversion in the no-error case.}
 \label{table:backwardsyndrome}       
\begin{tabular}{C{1.2cm}C{1.2cm}C{1.2cm}C{3.5cm}C{1.2cm}C{1.2cm}C{1.2cm}C{4cm}}
\hline\noalign{\smallskip}
$S'_1$ & $S'_2$ & $S'_3$ & operation & $S'_1$ & $S'_2$ & $S'_3$ & operation\\
\noalign{\smallskip}\hline\noalign{\smallskip}
0 & 0 & 0 & I & 1 & 0 & 0 & $Z_{2}Z_{3}Z_{10}Z_{11}$\\
0 & 0 & 1 & $Z_{3}Z_{7}Z_{11}Z_{15}$ & 1 & 0 & 1 & $Z_{2}Z_{7}Z_{10}Z_{15}$\\
0 & 1 & 0 & $Z_{1}Z_{3}Z_{9}Z_{11}$ & 1 & 1 & 0 & $Z_{1}Z_{2}Z_{9}Z_{10}$\\
0 & 1 & 1 & $Z_{1}Z_{7}Z_{9}Z_{15}$ & 1 & 1 & 1 & $Z_{1}Z_{2}Z_{3}Z_{7}Z_{9}Z_{10}Z_{11}Z_{15}$\\
\noalign{\smallskip}\hline
\end{tabular}
\end{table}

We now consider random single-qubit errors and their influence on the syndromes of Table~\ref{table:backwardsyndrome}
using an approach similar to that used in \S\ref{sec:convertSteaneRMQC}.
First we measure the~$Z$-type syndromes of Table~\ref{table:errorsyndrome} to deduce the single-qubit~$X$ error on qubit~$i$,
written as $X_i$.
This single-qubit error has no influence on the syndromes of Table~\ref{table:backwardsyndrome}
so we need only add the $X_i$ operation in the correction process.
For the~$X$-type syndromes,
as we already have the syndromes $S'_1$, $S'_2$ and $S'_3$,
similar as in \S\ref{sec:convertSteaneRMQC}, by generalizing Eq.~(\ref{eq:S141ZM}),
we obtain the simplifying assignments
\begin{align}
	\mathbb{M}\left(\bar{G}(1,4)_1^{X}\right)\to&S'_4=\mathbb{M}\left(X_9X_{11}X_{13}X_{15}\right),
		\nonumber\\
	\mathbb{M}\left(\bar{G}(1,4)_2^{X}\right)\to&S'_5=\mathbb{M}\left(X_{10}X_{11}X_{14}X_{15}\right),
		\nonumber\\
	\mathbb{M}\left(\bar{G}(1,4)_3^{X}\right)\to&S'_6=\mathbb{M}\left(X_{12}X_{13}X_{14}X_{15}\right),
        \nonumber\\
	\mathbb{M}\left(\bar{G}(1,4)_4^{X}\right)\to&S'_7=\mathbb{M}\left(X_{8}X_{9}X_{10}X_{11}\right).
\end{align}
The relations between these~$X$-type syndromes and $S(1,4)_{1,2,3,4}^{X}$
are
\begin{align}
\label{eq:backwardsynrelation}
 &S(1,4)_1^X=S'_1\oplus S'_4,~~~ S(1,4)_2^X=S'_2\oplus S'_5,\notag\\
 &S(1,4)_3^X=S'_3\oplus S'_6,~~~ S(1,4)_4^X=S'_6\oplus S'_7.
\end{align}

We use the syndrome set $\{S'_1,S'_2,\dots,S'_7\}$
and the relations in Eq.~(\ref{eq:backwardsynrelation}) as well as Table~\ref{table:errorsyndrome} to deduce the single-qubit~$Z$ error denoted $Z_j$.
Following we modify the first three syndromes according to
\begin{equation}
\label{eq:backwardsyndromecorrection}
	X_j\in \bar{G}(1,3)_n^X\implies S'_n=S'_n\oplus1
	(n=1,2,3).
\end{equation}
Now we use the fixed syndromes $S'_{1,2,3}$ and Table~\ref{table:backwardsyndrome} to get the operations written as $Z_\text{sub}$, the total operations we need to perform in the correction process is $Z_\text{sub}Z_jX_i$, by these operations we get the 15-qubit ERMQC. Then we can discard the last 8 qubits to get the Steane code.

The backward conversion circuit is shown in Fig.~2,
which shows that we need to measure four weight-8 Z type stabilizers and seven weight-4 X type stabilizers.
To guaranty fault tolerance,
the method used in \S\ref{sec:convertSteaneRMQC} to make fault-tolerant measurements should also be applied here.
The probability of obtaining the correct Steane code without
applying the fixing operation is also~$1/8$ for the no-error case as is the case for forward conversion.
Thus, with probability $1/8$, the procedure terminates here with no error.
Otherwise,
in order to obtain the correct Steane code without error,
then,
after the correction process,
we undertake another round of syndrome measurements;
this additional round needs six ancillary qubits to measure the stabilizers of~$\bar{G}(1,3)^X$ and~$\bar{G}(1,3)^Z$ to determine whether an error happened during the correction process.

If an error is detected,
we repeat the correction and syndrome measurement one more time.
Then, because we allow only one error, the procedure is complete.
As in \S\ref{sec:convertSteaneRMQC}, if we only want to fulfill fault-tolerant conversion, the ancillary qubits 8-11 in Fig.~2, which are used to diagnose the single X errors,
can be omitted.
Thus we need only measure seven weight-4 X type stabilizers.
\begin{figure}
\label{fig:backwardconversion}
\begin{center}

\includegraphics[width=0.8\textwidth]{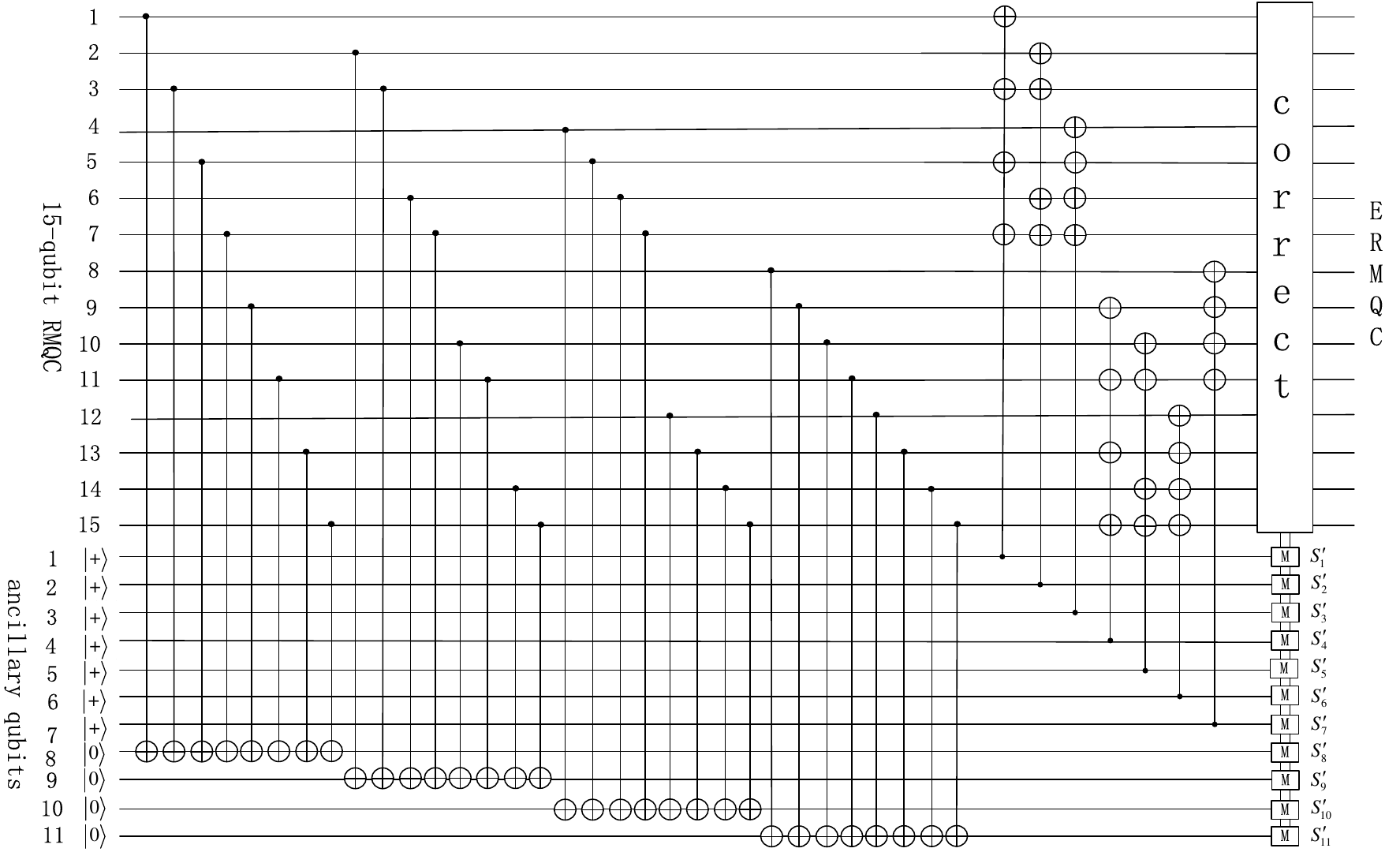}

\caption{%
Conversion circuit from 15-qubit RMQC to Steane code.
The ancillary qubits 1--11 are used to measure the syndromes
 $S'_1=\mathbb{M}\left(X_1X_3X_5X_{7}\right)$, $S'_2=\mathbb{M}\left(X_2X_3X_{6}X_{7}\right)$, $S'_3=\mathbb{M}\left(X_4X_5X_{6}X_{7}\right)$, $S'_4=\mathbb{M}\left(X_9X_{11}X_{13}X_{15}\right)$, $S'_5=\mathbb{M}\left(X_{10}X_{11}X_{14}X_{15}\right)$, $S'_6=\mathbb{M}\left(X_{12}X_{13}X_{14}X_{15}\right)$, $S'_7=\mathbb{M}\left(X_8X_{9}X_{10}X_{11}\right)$ , $S'_8=\mathbb{M}\left(Z_1Z_3Z_5Z_7Z_9Z_{11}Z_{13}Z_{15}\right)$, $S'_9=\mathbb{M}\left(Z_2Z_3Z_6Z_7Z_{10}Z_{11}Z_{14}Z_{15}\right)$, $S'_{10}=\mathbb{M}\left(Z_4Z_5Z_6Z_7Z_{12}Z_{13}Z_{14}Z_{15}\right)$, $S'_{11}=\mathbb{M}\left(Z_8Z_{9}Z_{10}Z_{11}Z_{12}Z_{13}Z_{14}Z_{15}\right)$.
At the end of the circuit we get the 15-qubit ERMQC, and then,
by deleting the last 8 qubits, we obtain Steane code.}
\end{center}
\end{figure}
\section{Conversion between adjacent RMQCs}
\label{sec:convertmtomplus1}
As the recursive relation~(\ref{eq:Gm1})
exists for all adjacent RMQCs,
denoted~$\overline{\textsf{RM}}(1,m)$ and~$\overline{\textsf{RM}}\left(1,m+1\right)$,
our method is suitable for all conversions between these adjacent codes.
Now we explain how to convert between adjacent RMQCs and the resources needed.

Beginning with the case $m=3$ and Eq.~(\ref{eq:Gm1}),
we obtain the generator matrix~$\bar{G}(1,m)$ and~$\bar{G}\left(1,m+1\right)$ for $\overline{\textsf{RM}}(1,m)$ and $\overline{\textsf{RM}}\left(1,m+1\right)$.
Accordingly we can define a matrix
\begin{equation}
	\tilde{H}\left(1,m+1\right)
		:=  \begin{bmatrix}
         \multicolumn{2}{c}{\tilde{H}\left(1,m+1\right)_{1}}\\
         \multicolumn{2}{c}{\vdots}\\
         \multicolumn{2}{c}{\tilde{H}\left(1,m+1\right)_{m}}\\
            \bar{G}(1,m)~~&\bm{\bar{0}}_{2^m}\\
               \tilde{H}(1,m)~~&\bm{\bar{0}}_{2^m}\\
               \bm{\bar{0}}_{2^m}~~&\tilde{H}(1,m)\\
               \end{bmatrix},
\end{equation}
from $m=4$
where
\begin{equation}
	\tilde{H}\left(1,m+1\right)_{1}
		=\bm{1}\left(1,3\right),
	\tilde{H}\left(1,m+1\right)_{2}
		=\bm{1}\left(2,3\right)
\end{equation}
and
\begin{equation}
	\tilde{H}\left(1,m+1\right)_{n}
		=\bm{1}\left(3,3+2^{n-1}\right),
		\;3\leq n\leq m,
\end{equation}
and $\bm{1}\left(x,y\right)$ means a vector is~$1$ at the positions of $x,y, x+2^m, y+2^m$, and~$0$ for the other elements.
We can verify
\begin{equation}
	\tilde{H}\left(1,m+1\right)\bar{G}\left(1,m+1\right)^T=0
\end{equation}
and the parity-check matrix for~$\overline{\textsf{RM}}(1,m+1)$,
namely,
$\bar{H}\left(1,m+1\right)$,
is the vertical concatenation of
$\bar{G}\left(1,m+1\right)$ and~$\tilde{H}\left(1,m+1\right)$.
Thus, the stabilizer group for $\overline{\textsf{RM}}\left(1,m+1\right)$ can be written as
\begin{equation}
\label{eq:SRM+1}
	\bm{S}\left(1,m+1\right)
		=\left\langle \bar{G}\left(1,m+1\right)^X, \bar{G}\left(1,m+1\right)^Z, \tilde{H}\left(1,m+1\right)^Z\right\rangle.
\end{equation}

The ERMQC for conversion from~$m$ to $m+1$ can be written as
\begin{equation}
	\ket{\phi}_{m\backsim m+1}
		=1/\sqrt{2}\left(\alpha\ket{\bar{0}}_{m}+\beta\ket{\bar{1}}_{m}\right)
			\left(\ket{0}\ket{\bar{0}}_{m}+\ket{1}\ket{\bar{1}}_{m}\right).
\end{equation}
The stabilizer group for this code can be written as
 \begin{align}
 \label{eq:SEXRM+1}
\bm{S}_{m\backsim m+1}=\Big\langle &\left(\bar{G}(1,m)\otimes\bm{\bar{0}}_{2^m}\right)^X,\bar{G}\left(1,m+1\right)^X, \left(\bar{G}(1,m)\otimes\bm{\bar{0}}_{2^m}\right)^Z,\notag  \\
&\bar{G}\left(1,m+1\right)^Z,\left(\tilde{H}(1,m)\otimes\bm{\bar{0}}_{2^m}\right)^Z,\left(\bm{\bar{0}}_{2^m}\otimes \tilde{H}(1,m)\right)^Z
  \Big\rangle.
\end{align}
Both the RMQC and ERMQC correspond to a subsystem code with stabilizer group of
\begin{align}
\bm{S}^\text{sub}_{m\backsim m+1}=\Big\langle &\bar{G}\left(1,m+1\right)^X,\bar{G}\left(1,m+1\right)^Z,\left(\bar{G}(1,m)\otimes\bm{\bar{0}}_{2^m}\right)^Z,\notag \\
&\left(\tilde{H}(1,m)\otimes\bm{\bar{0}}_{2^m}\right)^Z,\left(\bm{\bar{0}}_{2^m}\otimes \tilde{H}(1,m)\right)^Z\Big\rangle
\end{align}
with different gauge operators so they can convert to each other by gauge fixing.

For forward conversion,
which converts from $\overline{\textsf{RM}}(1,m)$ to $\overline{\textsf{RM}}\left(1,m+1\right)$, we compare Eq.~(\ref{eq:SEXRM+1}) with Eq.~(\ref{eq:SRM+1})
to discover that the first~$m$ stabilizers of $\tilde{H}\left(1,m+1\right)^Z$ may not be satisfied.
Thus, in order to convert forward,
we need only measure these~$m$ stabilizers and choose the corresponding operations to correct the collapsed state to $\overline{\textsf{RM}}\left(1,m+1\right)$. These $X$-type operations can be chosen similar as in \S\ref{sec:convertSteaneRMQC}, which commute with all the other stabilizers except for the stabilizers which were measured to be~$1$.

For backward conversion,
which converts from $\overline{\textsf{RM}}\left(1,m+1\right)$ to $\overline{\textsf{RM}}(1,m)$,
we discover that the stabilizers
	$\left(\bar{G}(1,m)\otimes\bm{\bar{0}}_{2^m}\right)^X$
of ERMQC may not be satisfied by comparing Eq.~(\ref{eq:SRM+1}) with Eq.~(\ref{eq:SEXRM+1}).
Thus, we need~$m$
ancillary qubits to measure these stabilizers and do corresponding operations to get the ERMQC.  Finally we drop the last $2^m$ qubits to obtain $\overline{\textsf{RM}}(1,m)$.

Consider the random single-qubit errors, for both directions of conversion, we need also measure the stabilizers of $\bar{G}\left(1,m+1\right)^X$ and $\bar{G}\left(1,m+1\right)^Z$ to distinguish the single-qubit errors, and fix the first~$m$ syndromes using the relations in Eqs.~(\ref{eq:syndromecorrection}) and~(\ref{eq:backwardsyndromecorrection}) with $n=1,2,\ldots,m$ respectively.
Then, by using the fixed syndromes,
we choose the correct fixing operations and perform error correction and fixing operations in one single step.

Following the above description,
we need to separately  measure $3m+2$ stabilizers for both forward and backward conversion.
For forward conversion we need to measure separately~$m+1$ weight-$2^{m}$ X, Z type stabilizers and~$m$ weight-4 Z type stabilizers, and,
for backward conversion,
we need to measure separately $m+1$ weight-$2^{m}$ X, Z type stabilizers and~$m$ weight-$2^{m-1}$ X type stabilizers. Whereas,
in Ref.~\cite{Anderson2014},
they measure all the $2^{m+1}-2$ stabilizers and refer to $2^{m+1}-m-2$ syndromes to do error correction. For the case $m=3$, the number of useful syndromes is the same, but for larger~$m$, our method reduces the number exponentially with~$m$.

Using the simplified method shown in Eq.~(\ref{eq:S141ZM}), the resources can be further reduced.
For forward conversion we need to measure $m+1$ weight-$2^{m}$ X type stabilizers, $m$ weight-4 Z type stabilizers, $m$ weight-$(2^{m}-4)$ Z type stabilizers plus one weight-$2^{m}$ Z type stabilizer; for the backward conversion we need to measure $m+1$ weight-$2^{m}$ Z type stabilizers and $2m+1$ weight-$2^{m-1}$ X type stabilizers. For the case $m=3$, if we divide the only weight-$2^{m}$ Z type stabilizer into two weight-4 stabilizers, the number of syndromes is 12 and the total weights of the syndromes are 64 and 60 respectively for forward and backward conversion, the same as shown in Figs.~1 and~2. If we only want to fulfill fault-tolerant conversion, the $m+1$ X type stabilizer measurements for the forward conversion and the $m+1$ Z type stabilizer measurements for the backward conversion can be omitted, and the number of stabilizers needed to be measured is reduced to $2m+1$.

\section{Simulation and cost analysis}
\label{sec:simulationandcost}

We simulate both directions of conversion using MATLAB$^{\tiny{\textregistered}}$ for the case of $m=3$ and $m=4$, where we introduce random single-qubit errors. The $X$-type stabilizer measurements are realized by first making a transversal~$\bar{H}$,
defined in Eq.~(\ref{eq:barHn}),
to the code, which converts an~$X$ measurement to a~$Z$ measurement and vice versa. Thus, we can measure the corresponding $Z$-type stabilizers in the dual code.
Then we use the transversal~$\bar{H}$ to convert back to the original code space.

The simulation process is depicted in Fig.~3.
The computer programme is used to verify that,
for any single-qubit errors in the input,
our single-step error-correction and code-conversion process is performing correctly.
The single-step process includes first combining the syndromes according to Eqs.~(\ref{eq:S141ZM}), (\ref{eq:S1423ZM}) and~(\ref{eq:backwardsynrelation}).
Then,
based on these syndromes,
we diagnose the single-qubit errors and use relations~(\ref{eq:syndromecorrection}) and~(\ref{eq:backwardsyndromecorrection}) to fix the first~$m$ syndromes and choose the corresponding operations. Then we implement the operations and compare the actual output with the ideal output to check whether the circuit and the single-step process is correct.

The result shows that,
if we use all $3m+2$ syndromes,
we obtain the correct output.
If we only use $2m+1$ syndromes,
we obtain the output with at most only one error.
This error does not propagate during conversion as it is fault-tolerant.

\begin{figure}
\begin{center}
 \includegraphics[width=\textwidth]{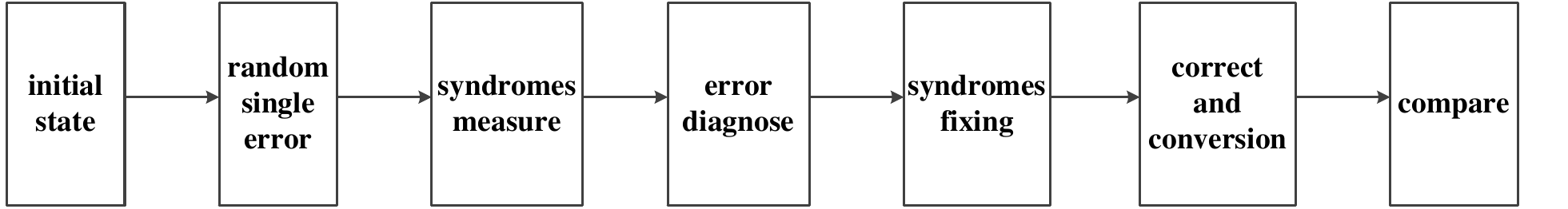}
\caption{%
The simulation process to verify our single-step error-correction and code-conversion process.
 }
\label{fig:simulationprocess}
\end{center}
\end{figure}

In order to see the huge decrease in requisite resources,
we compare the average resource requirement for three methods to realize the fault-tolerant logical $\bar{T}$ gate for the Steane code. These three methods are
the standard method~\cite{Nielsen2000,Fowler2011,Bremner2011}, the ADP14 method~\cite{Anderson2014} and our simplified method.
We assess using the average cost employed by Choi~\cite{Choi2015} which includes our previous counting of \textsc{CNOT} gates but also incorporates all gates, measurements and error rate.
Specifically, we apply Choi's average-cost assessment~\cite{Choi2015} to the standard method. The average cost for ADP14 to realize $\bar{T}$ is shown in Table~\ref{table:costADP}, the average cost for our method to realize $\bar{T}$ is shown in Table~\ref{table:costourmethod}.

In our calculation, we treat all physical gates and single-qubit measurements as having fixed unit cost.
We treat the error rates of the qubit, gate and measurement as all being equal to $\epsilon\in(10^{-6}-10^{-3})$~\cite{Steane2003}.
The calculation results for these three methods are shown in Table~\ref{table:averagecost}.

\begin{table}
\caption{ Cost(ADP14), where Cost(entangleS) can be found in Table~26 in~\cite{Choi2015} plus the cost for $\ket0_{Steane}$, AvgCost($S_{i,k}$) can be found in Table~12 in~\cite{Choi2015}.}
\label{table:costADP}
\begin{tabular}{C{3.5cm}C{7cm}C{5.5cm}}
\hline\noalign{\smallskip}
Cost or probability & Value or reference & Explanation \\
\noalign{\smallskip}\hline\noalign{\smallskip}
Cost(ancillary) & Cost(entangleS) & cost for the input\\
Cost($QEC_{RM}$) & $8\times \operatorname{AvgCost}(S_{i,8}$)+$6\times \operatorname{AvgCost}(S_{i,4}$)  & 14 stabilizer measurements\\
Cost(fix operation) & $0.875\times 4\times \operatorname{Cost}(X)$  & fix operation, Table~\ref{table:syndrome}\\
Cost($\bar{T}$) & $15\times \operatorname{Cost}(T)$  & Transversal T on RM code \\
Cost($QEC_{RM}$) & $8\times ¡¢\operatorname{AvgCost}(S_{i,8}$)+$ 6\times \operatorname{AvgCost}(S_{i,4}$)  & 14 stabilizer measurements\\
Cost(fix operation) & $0.75\times 4\times \operatorname{Cost}(Z)+0.125\times8\times \operatorname{Cost}(Z)$  & fix operation, Table~\ref{table:backwardsyndrome}\\
Average cost & Sum of all costs &  \\
\hline\noalign{\smallskip}
\end{tabular}
\end{table}

\begin{table}
\caption{ Cost(our method), where Cost(entangleS) can be found in Table~26 in~\cite{Choi2015} plus the cost for $\ket0_{Steane}$, AvgCost($S_{i,k}$) can be found in Table~12 in~\cite{Choi2015}. }
\label{table:costourmethod}
\begin{tabular}{C{3.5cm}C{7cm}C{5.5cm}}
\hline\noalign{\smallskip}
Cost or probability & Value or reference & Explanation \\
\noalign{\smallskip}\hline\noalign{\smallskip}
Cost(ancillary) & Cost(entangleS) & cost for the input\\
Cost($QEC_{RM}$) & $8\times \operatorname{AvgCost}(S_{i,4}$)  & 8 stabilizer measurements\\
Cost(fix operation) & $0.875\times4\times \operatorname{Cost}(X)$  & fix operation, Table~\ref{table:syndrome}\\
Cost($\bar{T}$) & $15\times \operatorname{Cost}(T)$  & Transversal T on RM code \\
Cost($QEC_{RM}$) & $ 7\times \operatorname{AvgCost}(S_{i,4}$)  & 7 stabilizer measurements\\
Cost(fix operation) & $ 0.75\times 4\times \operatorname{Cost}(Z)+0.125\times8\times \operatorname{Cost}(Z)$  & fix operation, Table~\ref{table:backwardsyndrome}\\
Average cost & Sum of all costs &  \\
\hline\noalign{\smallskip}
\end{tabular}
\end{table}

\begin{table}
\caption{ Average cost of fault-tolerant logical $\bar{T}$ for the standard method, ADP14 method and our method}
\label{table:averagecost}
\begin{tabular}{C{3cm}C{5cm}C{3cm}C{3cm}}
\hline\noalign{\smallskip}
Error rate ($\epsilon$) & standard method & ADP14 & our method\\
\noalign{\smallskip}\hline\noalign{\smallskip}
0.000001 & 932&4009 & 1882 \\
0.00001 & 932 &4010& 1883 \\
0.0001 & 937  &4020& 1887\\
0.001 & 981 &4127 & 1928\\
\hline\noalign{\smallskip}
\end{tabular}
\end{table}

From these results,
we see that our method reduces the resource overhead significantly compared with ADP14.
The main reason for this cost reduction is that we only measure eight weight-4 syndromes for the forward conversion and seven weight-4 syndromes for the backward conversion,
whereas ADP14 requires measurement of all eight weight-8 syndromes and six weight-4 syndromes. For larger $m$, the advantage of our method is evident as the syndromes are exponentially reduced from $2^{m+1}-2$ to $2m+1$.

Compared with the standard method for fault-tolerantly effecting a \emph{single} $\bar{T}$ gate,
our method needs approximately twice the resource of it,
as we first need to convert to the 15-qubit RMQC and then convert back to the Steane code after performing the transversal $\bar{T}$ gate.
This double overhead is not a problem, though.
We do not expect to convert back and forth between codes for each execution of $\bar{T}$.
Once in the 15-qubit RMQC,
we can execute many gates as long as the sequence is not interrupted by the $\bar{H}$ gate,
which is not transversal in that code.
Quantum-code conversion is valuable precisely for this reason:
order the operations in the circuit so as many gates as possible can be performed transversally in one code before converting to the other code for a sequence of gates that can be performed transversally in that code. Moreover, the conversion between the two codes is also extremely important because we could convert the 15-qubit RMQC to the smaller Steane code so as to operate most of the gates more efficiently.

\section{Conclusions}
\label{sec:conclusion}

We use the special case of converting from the Steane code to the 15-qubit RMQC as an example to explain the simplified code-conversion method,
in which the novel single-step process can diagnose the single qubit errors and also consider their influence on the gauge operators so as to fulfill
fault-tolerant conversion. The required resource is further decreased by the relations between gauge operators and stabilizers.

We extend this special case
to the general case of adjacent RMQCs, where the required number of stabilizer measurements is decreased to $2m+1$ from $2^{m+1}-2$.
In particular we provide explicit mathematical expression for all stabilizers that need to be measured for conversion,
and we discuss all requisite ancillary qubits and stabilizers weights.

Conversion between $\overline{\textsf{RM}}(1,3)$ and $\overline{\textsf{RM}}(1,4)$
enables a universal transversal set of gates.
Furthermore we can also convert between any adjacent order-one RMQCs so to non-adjacent RMQCs conversion, which enables all gates of the type
$\sqrt[2^{m-4}]{T}$
to be transversal for $m\geq4$~\cite{ZCC11}.

\begin{acknowledgements}
DXQ, LLZ and CXP acknowledge the support from National Natural Science Foundation of China No.\ 61372076 and No.\ 61301171, the 111 Project (No.\ B08038), the China Scholarship Council (No.\ 201506965068); BCS appreciates financial support from the Institute for Quantum Information and Matter, an NSF Physics Frontiers Center (NSF Grant PHY-1125565) with support of the Gordon and Betty Moore Foundation (GBMF-2644),
and financial support NSERC,  AITF, China's 1000 Talent Plan
and National Natural Science Foundation of China No.\ 11675164.
\end{acknowledgements}

\bibliography{qcodingwithoutdoi}
\end{document}